\documentclass[preprint,12pt]{elsarticle}
\usepackage{amsmath}
\usepackage{amsfonts}
\usepackage{amssymb}
\usepackage[utf8]{inputenc}
\usepackage[T2A]{fontenc}
\usepackage{bm}
\usepackage{graphicx}

\begin{document}

\title{General solution of the Dirac equation  \\
for quasi-two-dimensional electrons  }

\author{A.A. Eremko$^1$, L.S. Brizhik$^1$, V.M. Loktev$^{1,2}$}
\address{$^1$ Bogolyubov Institute for Theoretical Physics, \\
Metrologichna Str., 14-b,  Kyiv, 03680, Ukraine \\
$^2$ National Technical University of Ukraine "KPI", \\
Peremohy av., 37, Kyiv, 03056,  Ukraine}

\begin{abstract}
The general solution of the Dirac equation for quasi-two-dimensional electrons confined in an asymmetric quantum well, is found. The energy spectrum of such a system is exactly calculated using special unitary transformation and shown to depend on the electron spin polarization. The general solution, being the only one,  contains free parameters,  whose variation continuously transforms one known particular solution into another. As an example, two different cases are considered in detailL:  electron in a deep and in a strongly asymmetric shallow quantum well. The effective mass renormalized
 by relativistic corrections and Bychkov-Rashba coefficients are analytically obtained for both cases. The general solution allows - independently on the existence of the spin invariants - to establish conditions at which  a specific (accompanied or non-accompanied by Rashba splitting) spin state can be realized. In principle, this opens new possibilities of the spin degree of freedom control in spintronics via synthesis of heteroctructures of the desirable properties.
\end{abstract}

\maketitle


\section{Introduction}

Study of the spin degree of freedom is an important problem not only for fundamental physics, but for practical applications also, in view of the last advances of nanotechnologies and promising perspectives of the new branch of electronics called  spintronics. Of the special interest is spin behaviour in quasi-two-dimensional structures, such as layered semiconducting heterostructures, quantum wells (QW) and surface states of  compounds  widely used in modern micro- and nanoelectronics. In these systems there exist such phenomena as band spin splitting, or Rashba effect \cite{Rashba1, Rashba2}, spin Hall effect etc.  These phenomena are caused by spin-orbit interaction (SOI), which is manifested when particles propagate in an inhomogeneous  potential.  Therefore, the detailed study of the spin degree of freedom, the very existence of which  is the direct consequence of the Dirac equation (DE), is important.

Many problems of solid state electronics are usually studied within the Schr\"odinger equation (SE) taking into account SOI which appears in the Hamiltonian as the relativistic Thomas correction after the transition in DE to the non-relativistic limit  \cite{Bete, LandauIV, Davydov}. This correction takes the simplest form of the SOI operator in the case of a spherically symmetric field, and the form of Rashba SOI for two-dimensional (2D) carriers in the field of a plane well.   

It is generally assumed that SE with SOI operator describes all possible cases of entangling of coordinate and spin variables. Worth noting that DE does not include a separate term responsible for SOI.   Its appearance in the nonrelativistic approximation reflects the fact of the hidden coupling between the particle's degrees of freedom in the DE. The corresponding relation results from the fact that spatial and spin coordinates are not separated in the DE. This is very clear if one searches the solution of the DE for a free particle in the system of eigen functions that is consistent not with the momentum operator, but with the operator of the total angular momentum.   

To study how complete is the description of the spin degree of freedom within the nonrelativistic approximation, the solution of the DE for electrons in the field of the plane QW was recently analysed in \cite{arXiv} with the subsequent consideration of nonrelativistic energies.  It has been found, somehow unexpected, that there are \textit{four} independent solutions which correspond to also four different eigen spin states with different spin polarisations -- in other words, with different axes of spin momentum quantization -- and with different characters of SOI manifestation. Nonetheless, from the mathematical point of view there is nothing strange in this, according to the theory of linear differential equations.

Indeed, the DE is, in fact, the system of four equations for the components 
 $ \psi_{1}, \ldots , \psi_{4}  $  of the Dirac bispinor. In the external potential which varies in one direction only (for instance, along $ z $-axis), when the space in the two other directions is homogeneous, particle states can be characterized by certain value of the 2D momentum  $ \mathbf{k}_{\bot} = \mathbf{e}_{x} k_{x} + \mathbf{e}_{y} k_{y} $. Then the dependence of the wavefunction on the spatial coordinates  $ x $ and $ y $ is determined by the standard exponential term  $ \exp i \left( k_{x} x + k_{y} y \right)  $, and DE transforms into the system of four ordinary differential equations for functions  $ \psi_{j}(z) $ ($ j=1,\ldots ,4 $) of one variable. In this case one can find exactly four linear mutually independent  sets of functions   $ \psi_{\nu ,1}, \ldots , \psi_{\nu ,4} $ ($ \nu = 1, \ldots , 4 $), which represent the solution of the latter system of equations \cite{Kamke}. Such linearly independent solutions form \textit{the fundamental system} of solutions. For the system of solutions found in  \cite{arXiv}, the Wronskian is nonvanishing,  $ \textrm{Det} \vert \psi_{\nu,j}(z) \vert \neq 0 $, what unambiguously witnesses that the solutions are linearly independent. Thus, the solutions found in  \cite{arXiv} represent the fundamental system of solutions of the DE. Any  linear combination of them with arbitrary coefficients is also the solution of the DE  \cite{Kamke}. Such a solution of the system of equations is called \textit{general}, and, naturally, should  contain some free parameters, whose variation continuously transforms one solution into another.  

In view of the fact that different solutions correspond to different spin states, such a  variation of the free parameters means an arbitrary rotation of the spin polarization, which in its turn  leads to the appearance of SOI different for each state.  This is why it is important to find namely the general solution, or in other words, the solution with free parameters, when considering the DE for the specific physical situation. 
Such a solution, being a unique one, independently on the existence of the spin invariants, would allow to determine conditions at which a specific spin state is realized, and, hopefully, would open new possibilities of the control of the spin degree of freedom. Below we consider quasi-2D electrons kept by an asymmetric QW, and find the general solution of the DE. The effect of their spin polarization on the energy spectrum is also analysed. 

\section{Dirac Equation}

 Let us consider the stationary DE for electrons in an external potential:
\begin{equation}
\label{DE}
\left[ c\bm{\hat{\alpha}}  \bm{\hat{p}}  + V(\mathbf{r})\hat{I} + \hat{\beta} m c^{2} \right] \Psi = E \Psi \,,
\end{equation}
where $ c $ is the light velocity, $ m $ is particle mass, $ \bm{\hat{p}} = -i\hbar \bm{\nabla} $ is momentum operator, $ \bm{\hat{\alpha}} = \sum_{j} \mathbf{e}_{j} \hat{\alpha}_{j} $, $ \hat{\alpha}_{j} $ ($ j=x,y,z $) and $ \hat{\beta} $ are Dirac matrices, $ \hat{I} $ is a unit matrix, $V(\mathbf{r})$ is an external potential, and $ \Psi (\mathbf{r}) $ is a bispinor which is a four-component function of spatial coordinates. Below, unlike in our previous paper \cite{arXiv}, we will not explore the fact that there are the spin operators which commute with the Hamiltonian, but will use another approach.

In layered structures the potential varies in one direction, only, so that $ V(\mathbf{r}) = V(z) $ with  $ z $-axis chosen perpendicular to the layers. In this case, as we have stressed above, the integrals of motion are the two projections of particle momentum in the $ x y $-plane, and  the state of a particle with the given value of a 2D wavevector $ \mathbf{k}_{\perp} $ is described by the wavefunction
\begin{equation}
\label{Psi_2D}
\Psi_{\mathbf{k_{\bot}}}(\mathbf{r}) = e^{i(k_{x} x + k_{y} y)} \varPsi (z) \, .
\end{equation}
Substituting function (\ref{Psi_2D}) into Eq. (\ref{DE}), one can obtain the following equation: 
\begin{equation}
\label{1D-DE} 
\left[  c\hat{\alpha}_{z} \hat{p}_z + \hbar c \mathbf{k}_{\bot} \bm{\hat{\alpha}} + V(z)\hat{I} + m c^{2} \hat{\beta} \right] \varPsi (z) = E \varPsi (z) 
\end{equation}
for the bispinor $ \varPsi (z) $. We will use the $ 2\times 2 $ block form of Dirac matrices in their standard representation. Then Eq.  (\ref{1D-DE}) can be rewritten as $ \hat{H} \varPsi = E \varPsi $, with the Hamiltonian function 
\begin{equation}
\label{H_1}
\hat{H} = \left( 
\begin{array}{cc}
\left( V + mc^{2} \right) \hat{I}_{2} & c \hat{\sigma}_{z} \hat{p}_z + c\hbar \bm{\hat{\sigma}} \mathbf{k}_{\perp} \\ c\hat{\sigma}_{z} \hat{p}_z + c\hbar \bm{\hat{\sigma}} \mathbf{k}_{\perp} & \left( V - mc^{2} \right) \hat{I}_{2}
\end{array} \right) .
\end{equation}
The bispinor $ \varPsi (z) = \left( \psi_{1} \: \psi_{2} \: \psi_{3} \: \psi_{4} \right)^{T} $ can be represented in the form
\begin{equation}
\label{bispinor}
\varPsi (z) = { \psi_{u} (z) \choose \psi_{d} (z) } , \quad \psi_{u} (z) = { \psi_{1} (z) \choose \psi_{2} (z) } , \quad \psi_{d} (z) = { \psi_{3} (z) \choose \psi_{4} (z) } ,
\end{equation}
where $ \psi_{u} $ and $ \psi_{d} $ are upper and lower spinors, respectively.  The operator  $ \hat{I}_{2} $ in Eq. (\ref{H_1}) is a two-row unity matrix, and  $ \hat{\sigma}_{j} $ ($ j=x,y,z $) are Pauli matrices. 

Matrix equatin (\ref{1D-DE}) takes the form of the system of equations  
\begin{equation}
\label{Eqs_2D}
\left\lbrace  
\begin{array}{c}
 c \hat{p}_z\psi_{1}  + \hbar c k_{\perp} e^{-i\varphi_{\bot}} \psi_{2} - \left[ mc^{2} - V(z) \right] \psi_{3} = E \psi_{3} , \\
- c \hat{p}_z \psi_{2}  + \hbar ck_{\perp} e^{i\varphi_{\perp}} \psi_{1} - \left[ mc^{2} - V(z) \right] \psi_{4} = E \psi_{4} , \\
 c \hat{p}_z \psi_{3} + \hbar c k_{\perp} e^{-i\varphi_{\perp}} \psi_{4} + \left[ mc^{2} + V(z) \right] \psi_{1} = E \psi_{1} , \\
-c \hat{p}_z  \psi_{4}  + \hbar c k_{\perp} e^{i\varphi_{\bot}} \psi_{3} + \left[ mc^{2} + V(z) \right] \psi_{2} = E \psi_{2} , 
\end{array}
\right. 
\end{equation}
in which we have chosen the polar coordinates for the components of a 2D wavevector, $ k_{x} = k_{\perp} \cos \varphi_{\bot} $, $ k_{y} = k_{\perp} \sin \varphi_{\bot} $, where $ k_{\perp} = \sqrt{k_{x}^{2} + k_{y}^{2}} $, $ \tan \varphi_{\bot} = k_{y}/k_{x} $.

In some cases  the integration of differential equations can be performed if one finds  \textit{integrable combinations} of the form 
\[
\frac{d}{dz} \Phi \left( z;\psi_{1},\psi_{2},\psi_{3},\psi_{4} \right) = 0 .
\]
Such a combination, leading to the equality 
$ \Phi \left( z; \psi_{1},\psi_{2},\psi_{3},\psi_{4} \right) = \mathrm{const} $, even  not being the only one,  is called   \textit{the integral} of the system of equations and allows to reduce the number of the searched functions. Let us show that the system of Eqs. (\ref{Eqs_2D}) possesses such an integral. For this we multiply the first equation by  $ \psi_{4} $, the second one by $ - \psi_{3} $,the third one by $ \psi_{2} $, and the fourth one by  $ - \psi_{1} $. Summing up the results, we obtain the equality 
\[
\hat{p}_z \left( \psi_{1} \psi_{4} + \psi_{2} \psi_{3} \right) = 2 \hbar k_{\perp} \left( e^{i\varphi_{\bot}} \psi_{1} \psi_{3} - e^{-i\varphi_{\bot}} \psi_{2}\psi_{4} \right) .
\]
Similarly, multiplying the first equation by  $ \exp (i\varphi_{\bot}) \psi_{1} $, the second one by $ -\exp (-i\varphi_{\bot}) \psi_{2} $, the third one by $ -\exp (i\varphi_{\bot}) \psi_{3} $ and the fourth one by $ \exp (-i\varphi_{\bot}) \psi_{4} $ with the following summing of the results, we obtain another relation
\[
\hat{p}_z \left[ e^{i\varphi_{\bot}} \left( \psi_{1}^{2} - \psi_{3}^{2} \right) +  e^{-i\varphi_{\bot}} \left( \psi_{2}^{2} - \psi_{4}^{2} \right) \right] = 4mc \left( e^{i\varphi_{\bot}} \psi_{1} \psi_{3} - e^{-i\varphi_{\bot}} \psi_{2}\psi_{4} \right) .
\]
Comparing the latter two equalities we come to the identity 
\[
\frac{d}{dz} \left\lbrace \hbar k_{\perp} \left[ e^{i\varphi_{\bot}} \left( \psi_{1}^{2} - \psi_{3}^{2} \right) + e^{-i\varphi_{\bot}} \left( \psi_{2}^{2} - \psi_{4}^{2} \right) \right] - 2mc \left( \psi_{1}\psi_{4} + \psi_{2}\psi_{3} \right) \right\rbrace = 0 ,
\]
from which it follows that the combination 
\begin{equation}
\label{rel1}
\frac{\hbar k_{\perp}}{mc} \left[ e^{i\varphi_{\bot}} \left( \psi_{1}^{2} - \psi_{3}^{2} \right) + e^{-i\varphi_{\bot}} \left( \psi_{2}^{2} - \psi_{4}^{2} \right) \right] - 2 \left( \psi_{1}\psi_{4} + \psi_{2}\psi_{3} \right) = 0 , 
\end{equation}
determines the first integral of the system  (\ref{Eqs_2D}). Since the equality  (\ref{rel1}) ought to be fulfilled at arbitrary  $ z $, the constant of integration has been chosen equal to zero, according to the well-known physical requirement that the wavefunction vanishes at infinity,  $ z = \pm \infty $.

The existence of this integral shows that the number of the searched functions can be reduced. In particular, in Ref.
\cite{arXiv} the initial system of equations (\ref{Eqs_2D}) has been reduced to the system of two equations using another fact. Namely, it has been used that there are spin operators which commute with the Dirac Hamiltonian, and, hence, have a common with it system of eigenfunctions.  Since the invariant combinations do not contain the  derivatives with respect to $ z $-coordinate, the corresponding solutions of the equations for the eigenvalues can be represented in the form of a linear combination of two unknown functions, with an arbitrary dependence of the bispinor components on the spatial coordinate. Here we note that substituting solutions, found in \cite{arXiv} into the relation  (\ref{rel1}),  the latter is identically satisfied. Below we will prove that a similar reduction of Eqs. (\ref{Eqs_2D}) to the system of two equations is valid also in the case, when one searches for the general solution of the DE.  

\section{Reduction of the Dirac Hamiltonian dimension}

To find this solution, let use the method of unitary transformations and represent the bispinor  $ \varPsi (z) $ in the following form: 
\begin{equation}
\label{PsiTrans}
\varPsi (z) = \hat{U} \tilde{\varPsi} (z) , 
\end{equation}
where
\begin{equation}
\label{U_trans}
\hat{U} = \frac{1}{\sqrt{2 \varepsilon \left( \mathbf{K}_{\perp} \right) \left[ \varepsilon \left( \mathbf{K}_{\perp} \right)  +  mc^{2} \right]  } } \left( \begin{array}{cc}
\left[ \varepsilon \left( \mathbf{K}_{\perp} \right) + mc^{2} \right] \hat{\omega}_{u} & - \hbar c \hat{\omega}_{u}  \bm{\hat{\sigma}} \mathbf{K}_{\perp} \\ \hbar c \hat{\omega}_{d} \bm{\hat{\sigma}} \mathbf{K}_{\perp} & \left[ \varepsilon \left( \mathbf{K}_{\perp} \right) + mc^{2} \right] \hat{\omega}_{d}
\end{array} \right)  
\end{equation}
is an operator of the unitary transformation, $ \hat{U}^{\dagger} \hat{U} = \hat{I} $, 
\begin{equation}
\label{U_pm}
\hat{\omega}_{u} = \left(  \begin{array}{cc} 
\delta & - \gamma^{\ast} \\ \gamma & \delta^{\ast} 
\end{array}  \right) , \quad \hat{\omega}_{d} = \left(  \begin{array}{cc} 
\delta & \gamma^{\ast} \\ - \gamma & \delta^{\ast} 
\end{array}  \right)
\end{equation}
are matrices of some arbitrary rotation of the spinors in a 3D space.
The Keyli-Klein parameters,  $ \delta $ and $ \gamma $, due to the normalization  $ \vert \delta \vert^{2} + \vert \gamma \vert^{2} = 1 $ contain three independent real parameters, which can be expressed as
\begin{equation}
\label{KK-Eil} 
\delta = e^{(i/2)\left( \Theta - \varphi \right) } \cos \frac{\theta}{2} \, , \quad \gamma =  e^{(i/2)\left( \Theta + \varphi \right) } \sin \frac{\theta}{2} \, .
\end{equation}
The following notations have been used in the transformation operator  (\ref{U_trans}): 
\begin{equation}
\label{epsil}
\varepsilon \left( \mathbf{K}_{\perp} \right) = \sqrt{m^{2}c^{4} + \hbar^{2} c^{2} K_{\perp}^{2}} , \quad K_{\perp}^{2} = K_{x}^{2} + K_{y}^{2},
\end{equation}
where $ \mathbf{K}_{\bot} = \mathbf{e}_{x} K_{x} + \mathbf{e}_{y} K_{y} $ is a 2D vector, whose components are connected with the components of the vector  $ \mathbf{k}_{\bot} $ 
%via angles $ \Theta , \varphi , \theta $.  
%The components of the vectors  $ \mathbf{k}_{\bot} $ and $ \mathbf{K}_{\bot} $ are connected 
by the {\it{non-unitary}} rotation 
\begin{equation}
\label{K_x-K_y}
\begin{array}{c}
K_{x} = \left( \sin \Theta \sin \varphi \cos \theta + \cos \Theta \cos \varphi \right) k_{x} - \left( \sin \Theta \cos \varphi \cos \theta - \cos \Theta \sin \varphi \right) k_{y} = \\
= k_{\bot} \left[ \sin \Theta \cos \theta \sin \left( \varphi - \varphi_{\bot} \right) + \cos \Theta \cos \left( \varphi - \varphi_{\bot} \right) \right] , \\
K_{y} = - \left( \cos \Theta \sin \varphi \cos \theta - \sin \Theta \cos \varphi \right) k_{x} +  \left( \cos \Theta \cos \varphi \cos \theta + \sin \Theta \sin \varphi \right) k_{y} =  \\
= k_{\bot} \left[ - \cos \Theta \cos \theta \sin \left( \varphi - \varphi_{\bot} \right) + \sin \Theta \cos \left( \varphi - \varphi_{\bot} \right) \right], 
\end{array}
\end{equation}
because, as it is easy to check,   $ \vert \mathbf{k}_{\bot} \vert \neq \vert \mathbf{K}_{\bot} \vert $. Let us note, that appeared above  angles $ \Theta , \varphi , \theta $ are free parameters. 

Performing the transformation (\ref{U_trans}) explicitly, one can obtain the equation  $ \hat{\tilde{H}} \tilde{\varPsi} = E \tilde{\varPsi} $ for the bispinor $ \tilde{\varPsi} (z) = \left( \tilde{\psi}_{1} \: \tilde{\psi}_{2} \: \tilde{\psi}_{3} \: \tilde{\psi}_{4} \right)^{T} $, where 
\begin{equation}
\label{H_tilde}
\hat{\tilde{H}} = \hat{U}^{\dagger} \hat{H} \hat{U} = \left( 
\begin{array}{cc}
\left[ V(z) + \varepsilon \left( \mathbf{K}_{\perp} \right) \right] \hat{I}_{2} &  c  \hat{\sigma}_{z} \hat{p}_z - i \hbar c q \hat{I}_{2} \\  c  \hat{\sigma}_{z} \hat{p}_z + i \hbar c q \hat{I}_{2} & \left[ V(z) - \varepsilon \left( \mathbf{K}_{\perp} \right] \right) \hat{I}_{2} 
\end{array} \right) 
\end{equation}
is the 'rotated' Hamilton matrix  (\ref{H_1}), the energy  $ \varepsilon \left( \mathbf{K}_{\perp} \right) $ is determined by the expression (\ref{epsil}) and the notation  
\begin{equation}
\label{q} 
q = \sin \theta \left( k_{x} \sin \varphi - k_{y} \cos \varphi \right) = k_{\perp} \sin \theta \sin  \left( \varphi - \varphi_{\bot} \right) \equiv k_{\bot} f(\varphi_{\bot}) 
\end{equation}
is introduced. This parameter, as it will be shown below, is connected with Rashba splitting. 

The matrix equation $ \hat{\tilde{H}} \tilde{\varPsi} = E \tilde{\varPsi} $ in the explicit form, as it is easy to see, is reduced to the uncoupled pairs for the upper ($ \tilde{\psi}_{1}, \tilde{\psi}_{3} $) and lower ($ \tilde{\psi}_{2} , \tilde{\psi}_{4} $) functions  of the corresponding spinors:
\begin{equation}
\label{tildaEqs1} 
\left\lbrace \begin{array}{c}
c \hat{p}_z \tilde{\psi}_{1} + i \hbar c q \tilde{\psi}_{1} = \left[ E - V(z) + \varepsilon \left( \mathbf{K}_{\perp} \right) \right] \tilde{\psi}_{3} ,  \\ 
 c \hat{p}_z \tilde{\psi}_{3} - i \hbar c q \tilde{\psi}_{3} = \left[ E - V(z) - \varepsilon \left( \mathbf{K}_{\perp} \right) \right] \tilde{\psi}_{1} , 
\end{array} 
\right. 
\end{equation}
\begin{equation}
\label{tildaEqs2} 
\left\lbrace  
\begin{array}{c}
- c \hat{p}_z \tilde{\psi}_{2} + i \hbar c q \tilde{\psi}_{2} = \left[ E - V(z) + \varepsilon \left( \mathbf{K}_{\perp} \right) \right] \tilde{\psi}_{4} ,  \\ 
- c \hat{p}_z \tilde{\psi}_{4} - i \hbar c q \tilde{\psi}_{4} = \left[ E - V(z) - \varepsilon \left( \mathbf{K}_{\perp} \right) \right] \tilde{\psi}_{2} , 
\end{array}
\right. 
\end{equation}
for which the definitions $ \tilde{\psi}_{u} = \left( \tilde{\psi}_{1} \: \tilde{\psi}_{2} \right)^{T} $ and $ \tilde{\psi}_{d} = \left( \tilde{\psi}_{3} \: \tilde{\psi}_{4} \right)^{T} $ remain valid.

 It is important to bear in mind that, despite of the dimension reduction of the DE system (\ref{Eqs_2D}), one can't consider the solutions of  systems (\ref{tildaEqs1}) and (\ref{tildaEqs2}) independently, because the functions  $ \tilde{\psi}_{j} $ ($ j=1,2,3,4 $) are the components of the initial Dirac bispinor. They have to satisfy the above found integral  (\ref{rel1}) of the system (\ref{Eqs_2D}). In the presentation  (\ref{U_trans})  this integral takes the form 
\begin{equation}
\label{rel1_tr}
\varepsilon \left( \mathbf{K}_{\perp} \right) \left( \tilde{\psi}_{1} \tilde{\psi}_{4} + \tilde{\psi}_{2} \tilde{\psi}_{3} \right) - i \hbar c q \left( \tilde{\psi}_{1} \tilde{\psi}_{2} + \tilde{\psi}_{3} \tilde{\psi}_{4} \right) = 0 .
\end{equation}

Meanwhile the systems of Eqs.  (\ref{tildaEqs1}) and (\ref{tildaEqs2}) are uncoupled and are not conjugate with one another in the terms of the theory of homogeneous  systems  of linear differential equations  \footnote{The systems of equations  $ du_{j}/dz = \sum_{p=1}^{n}f_{j,p}(z) u_{p} $ and $ dv_{j}/dz = - \sum_{p=1}^{n}f_{p,j}(z) v_{p} $ are called mutually conjugate. Solutions of one system are expressed through the solutions of the other one  \cite{Kamke}.}. Therefore, the condition  (\ref{rel1_tr}) can be fulfilled identically only in the case when one of the systems has a trivial solution.  

In the coordinate system related with the spin quantization axis,  particle  states are described by two orthogonal bispinors, such that in one of them the lower component, and in the other one the upper component are equal to zero. The quantum number $ \sigma $ is prescribed to these spinors and distinguishes them. It takes the discrete values $ \sigma = \pm 1 $, which correspond to the states with opposite directions of the particle eigen angular momentum (spin). 

Let us associate the spin quantization axis with the system of bispinor components  $ \tilde{\varPsi} (z) $ and write down the solutions of Eqs. (\ref{tildaEqs1})-(\ref{tildaEqs2}) for the two orthogonal bispinors, prescribing them the spin quantum  number which, as described above, takes two values: 
\[
\tilde{\varPsi}_{+} (z) = \left( \begin{array}{c} \tilde{\psi}_{u+} (z) \\ \tilde{\psi}_{d+} (z) \end{array} \right) = \left( \begin{array}{c} u_{+} (z) \\ 0 \\ v_{+} (z) \\ 0
\end{array} \right) , \qquad  \tilde{\varPsi}_{-} (z) = \left( \begin{array}{c} \tilde{\psi}_{u-} (z) \\ \tilde{\psi}_{d-} (z) \end{array} \right) = \left( \begin{array}{c} 0  \\ u_{-} (z) \\ 0  \\ v_{-} (z)
\end{array} \right) .
\]
In the general form this can be rewritten as
\begin{equation}
\label{sol_up-d}
\tilde{\varPsi}_{\sigma} (z) = { u_{\sigma} (z) \chi_{\sigma} \choose v_{\sigma} (z) \chi_{\sigma} } , \quad \chi_{+} = { 1 \choose 0 } , \; \chi_{-} = { 0 \choose 1 } .
\end{equation}
The bispinor with  $ \sigma = 1 $ corresponds to the trivial solution of Eqs. (\ref{tildaEqs2}), $ \tilde{\psi}_{2} = \tilde{\psi}_{4} = 0 $, while the spin quantum number $ \sigma = -1 $ corresponds to similar solution of Eqs. (\ref{tildaEqs1}), $ \tilde{\psi}_{1} = \tilde{\psi}_{3} = 0 $. The non-trivial solutions  can be represented as  $ \tilde{\psi}_{1} = u_{+} $, $ \tilde{\psi}_{3} = v_{+} $ for $ \sigma = 1 $, and $ \tilde{\psi}_{2} = u_{-} $, $ \tilde{\psi}_{4} = v_{-} $ for $ \sigma = -1 $. It is easy to be convinced that functions  $ u_{\sigma} $ and $ v_{\sigma} $ are the solutions of equations 
\begin{equation}
\label{psi-phi} 
\left\lbrace  
\begin{array}{c}
 \sigma c \hat{p}_z u_{\sigma} + i \hbar c q u_{\sigma} = \left[ E - V(z) + \varepsilon \left( \mathbf{K}_{\perp} \right) \right] v_{\sigma} , \\
 \sigma  c \hat{p}_z v_{\sigma} - i \hbar c q v_{\sigma} = \left[ E - V(z) - \varepsilon \left( \mathbf{K}_{\perp} \right) \right] u_{\sigma} . 
\end{array}
\right. 
\end{equation} 

Substituting functions  (\ref{sol_up-d}) in the transformation (\ref{PsiTrans}) and taking into account the operator (\ref{U_trans}), for the needed bispinor we obtain the expression (\ref{bispinor}), in which upper and lower spinors have the form
\begin{equation}
\label{psi_u} 
\tilde{\psi}_{u\sigma} \left( z \right) = \sqrt{\frac{\varepsilon\left(\mathbf{K_{\bot}} \right) +mc^{2} }{2\varepsilon\left(\mathbf{K_{\bot}} \right) }} \left[ u_{\sigma} (z) \tilde{\chi }_{\sigma} - \frac{\hbar c \left( K_{x} + i\sigma K_{y} \right) }{\varepsilon\left(\mathbf{K_{\bot}} \right) +mc^{2}} v_{\sigma}(z)  \tilde{\chi}_{-\sigma} \right] ,
\end{equation} 
\begin{equation}
\label{psi_d} 
\tilde{\psi}_{d\sigma} \left( z \right) = \sqrt{\frac{\varepsilon\left(\mathbf{K_{\bot}} \right) +mc^{2} }{2\varepsilon\left(\mathbf{K_{\bot}} \right) }} \left[ v_{\sigma} (z) \tilde{\eta}_{\sigma} + \frac{\hbar c \left( K_{x} + i\sigma K_{y} \right) }{\varepsilon\left(\mathbf{K_{\bot}} \right) +mc^{2}} u_{\sigma}(z) \tilde{\eta}_{-\sigma} \right] , 
\end{equation}
where we have taken into account the equality  $ \bm{\hat{\sigma}} \mathbf{K}_{\perp} \chi_{\sigma} = \left( K_{x} + i\sigma K_{y} \right)\chi_{-\sigma} $ and used the following notations: $ \tilde{\chi}_{\sigma} = \hat{\omega}_{u} \chi_{\sigma} $ with the explicit expressions given in Eq. (\ref{spinors}),  $ \tilde{\eta}_{\sigma} = \hat{\omega}_{d} \chi_{\sigma} $. Functions $ u_{\sigma} $ and $ v_{\sigma} $ satisfy the system of Eqs. (\ref{psi-phi});  hence, the bispinor  (\ref{bispinor}) with these spinors is the solution of the DE  (\ref{1D-DE}) and can be used to calculate the probabilities of  physical observables in the experimental data.  

As it has been mentioned above, in  \cite{arXiv} the DE was solved with the use of spin invariants, which are $ z $-components of the operators of electric,   $ \hat{\epsilon}_{z} $, and magnetic, $ \hat{\mu}_{z} $, spin polarizations, and two orthogonal, $ x $- and $ y $-components of the spatial part of four-dimensional spin pseudo-vector,   $ \bm{\hat{\mathcal{S}}} $, \cite{Sokolov}. Notice, the solutions, found in \cite{arXiv}, are partial cases of the general solution  obtained here, (\ref{psi_u})-(\ref{psi_d}), at certain values of the free parameters. In particular,  at $ \theta = \pi /2 $ and $ \varphi = \varphi_{\bot} \pm \pi/2 $ the bispinor  (\ref{bispinor}) is eigen bispinor of the operator $ \hat{\epsilon}_{z} $, at $ \theta = 0 $ and $ \varphi = \Theta $ it is eigen bispinor of the operator $ \hat{\mu}_{z} $,  at  $ \theta = \pi /2 $ and $ \varphi = \phi_{0} $ it is eigen bispinor of the operator $ \bm{\hat{\mathcal{S}}}\cdot \mathbf{e}_0 $, i.e., of the vector   $ \bm{\hat{\mathcal{S}}} $ as projection on some direction  $ \mathbf{e}_0 = \mathbf{e}_{x} \cos \phi_{0} + \mathbf{e}_{y} \sin \phi_{0} $, which in fact is arbitrary in $xy$-plane. Indeed, at these specific values of the parameters the transformation  (\ref{U_trans}) diagonalizes the corresponding operators, 
$ \hat{A} $ $\left(= \hat{\epsilon}_{z},\,\hat{\mu}_{z} \,    {\rm{or}} \, \bm{\hat{\mathcal{S}}}\cdot \mathbf{e}_0 \right) $ reducing them to the form  $ \hat{U}^{\dagger} \hat{A} \hat{U} = \lambda \hat{\Sigma}_{z} $, where 
\[
\hat{\Sigma}_{z} = \left( 
\begin{array}{cc}
\hat{\sigma}_{z} & 0 \\ 0 & \hat{\sigma}_{z} 
\end{array} \right) .
\]

Obviously, the bispinor  (\ref{sol_up-d}) are eigen bispinors of the operator  
\[
\hat{U}^{\dagger} \hat{A} \hat{U} \tilde{\varPsi}_{\sigma} (z) = \sigma \lambda \tilde{\varPsi}_{\sigma} (z) 
\]
with some eigen values  $ \lambda $. For the operator $ \hat{A} = \hat{\epsilon}_{z} $ one has to set   $ \theta = \pi /2 $, $ \varphi = \varphi_{\bot} \pm \pi/2 $ in Eq. (\ref{U_trans}), which gives  $ \lambda = \hbar k_{\perp} $. For the operator $ \hat{A} = \hat{\mu}_{z} $ at $ \theta = 0 $ and $ \varphi = \Theta $ we get $ \lambda = \sqrt{1 + \left( \hbar k_{\perp}/mc \right)^{2} } $, and for $ \hat{A} = \bm{\hat{\mathcal{S}}}  \mathbf{e}_0 $, setting $ \theta = \pi /2 $ and $ \varphi = \phi_{0} $, we get $ \lambda = \sqrt{1 + \left( \hbar k_{0}/mc \right)^{2} } $ with $ k_{0} = \mathbf{k}_{\perp}  \mathbf{e}_0 $. Therefore, at these certain values of the parameters, the bispinors   (\ref{PsiTrans}) are the eigen bispinors of the corresponding spin operators. 

In the case of the homogeneous space, $ V(z) = \mathrm{const} $, as it is shown in Appendix, the solution of the system (\ref{psi-phi}) transforms into the well-known result.

\section{Electron confined in a quantum well}

To illustrate the various possibilities of the general solution, let us consider the   states of quasi-2D electrons confined by the potential of the following form: 
\begin{equation}
\label{V_123}
V(z) = \left\lbrace \begin{array}{ll}
V_{L}, &   z< - d/2 , \\
V_{C}(z) = V_{C}- \mathcal{E} z, & -d/2 < z < d/2 , \\
V_{R}, &  z> d/2 ,
\end{array} \right. 
\end{equation}
which is formed by the QW of the width $ d $, setting coordinate system at its center,  $ z = 0 $.  Without loss of generality one can adopt that $ V_{L} \geqslant V_{R} > V_{C}(z) = -\mathcal{E} z $, i.e., we will set $ V_{C} = 0 $. This means that the space in $z$-direction is naturally split into the three regions: left,  $ L $, central, $ C $, and right, $ R $, ones in each of which the potential has different values $ V_{j} $ ($ j = L,C,R $), respectively. In the central region  the dependence on   $ z $ is determined by the difference of potential values on the boundaries of the QW, i.e., by $ \mathcal{E}d $. In each region the state of an electron with  the given energy is determined by its wavefunction,  
\begin{equation}
\label{Psi_123}
\Psi (x,y,z) = \left\lbrace \begin{array}{ll}
\Psi_{L}(x,y,z), &   z< -d/2 , \\
\Psi_{C}(x,y,z), &  -d/2 < z < d/2  , \\
\Psi_{R}(x,y,z), &  z> d/2  ,
\end{array} \right. 
\end{equation} 
where $ \Psi_{j}$ are the solutions of the DE in the corresponding regions,  $j=L,C,R $. At the boundaries $ z=-d/2 $ and $ z=d/2 $ the solutions should transform into one another continuously, which can be formulated as the boundary conditions for the bispinors $ \varPsi_{j} $, neglecting their exponential multiplier:
\begin{equation}
\label{bcond}
\varPsi_{L} \left( z=-\frac{d}{2}\right)  = \varPsi_{C} \left( z=-\frac{d}{2}\right) , \quad \varPsi_{C} \left( z=\frac{d}{2}\right)  = \varPsi_{R} \left( z=\frac{d}{2}\right) . 
\end{equation}
The equality of bispinors at the boundaries corresponds to the equalities of their four components. 

In the following we will be interested in nonrelativistic energies, actual in condensed state physics.  In this case the inequality $ V_{L} \ll mc^{2}$ is always met, and we can consider the bound states in QW with positive energies  
\begin{equation}
\label{E_nonrel}
E = mc^{2} + E_{b}(\mathbf{k}_{\perp}) , 
\end{equation}
when the energy of the 2D bands $ E_{b}(\mathbf{k}_{\perp}) $ falls in the range $ 0 < E_{b}(\mathbf{k}_{\perp}) < V_{R } $. 

For the solutions of the system (\ref{psi-phi}), which correspond to such energies, functions $ u_{\sigma} $ and $ v_{\sigma} $ are large and small components, respectively. The system of two linear equations of the first order can be reduced to one equation of the second order for the large component  $ u_{\sigma} (z) $, expressing the small component $ v_{\sigma} (z) $ at the given energy via  function $ u_{\sigma} $: 
\begin{equation}
\label{varphi}
v_{\sigma} (z) = - i \frac{\sigma \hbar c}{E+ \varepsilon (\mathbf{K}_{\perp}) - V_{j}} \left( \frac{du_{\sigma} (z) }{dz} - \sigma q u_{\sigma} (z) \right) .
\end{equation}

Multiplying the second equation in (\ref{psi-phi}) by $ E - V(z) + \varepsilon (\mathbf{K}_{\perp}) $, and taking into account the first one, we get 
\[
-\hbar^{2} c^{2} \frac{d^{2} u_{\sigma}}{dz^{2}} - i\sigma \hbar c \frac{dV(z)}{dz} v_{\sigma} = \left\lbrace \left( E - V(z) \right)^{2} - \left[ \varepsilon^{2}\left( \mathbf{K}_{\perp} \right) + \hbar^{2} c^{2} q^{2} \right] \right\rbrace u_{\sigma} .
\]
At the constant values of the potential in the regions $ L $ and $ R $, $ V_{j} = \textrm{const} $ ($ j = L,R $), the equation for $ u_{\sigma}(z) $ takes the form
\begin{equation}
\label{Eq_psi_LR}
- \hbar^{2} c^{2}\frac{d^{2} u_{\sigma}}{dz^{2}} = \left[ \left( E - V_{j} \right) ^{2} - \varepsilon^{2}\left( \mathbf{k}_{\perp} \right) \right] u_{\sigma} .
\end{equation}
Here we have taken into account the definition (\ref{epsil}) and equality $ K_{\perp}^{2} + q^{2} = k_{\perp}^{2} $, which follows from the explicit forms of the vector  $ \mathbf{K}_{\perp} $ and value $ q $ (see Eqs. (\ref{K_x-K_y}) and (\ref{q})). For the bound states with the energies, as described above, the multiplier $ [\left( E - V_{j} \right) ^{2} - \varepsilon^{2}\left( \mathbf{k}_{\perp} \right)] $ ($ j = L,R $) in the right hand side of Eq. (\ref{Eq_psi_LR}) is negative, i.e.,  $ \left[\cdots \right] < 0 $. Hence, beyond the region of the QW,  at $ \vert z \vert > d/2 $, the function $ u_{\sigma} $ satisfies the equation $ d^{2} u_{\sigma}/dz^{2} = \kappa^{2} u_{\sigma} $, whose solutions are exponentially increasing and decreasing functions.

The solution which vanishes at $ z \rightarrow -\infty $, has the physical meaning in the left region, while the solution vanishing at $ z \rightarrow \infty $, has the physical meaning in the right region:
\[
u_{L} = C_{L} \exp \left[ \kappa_{L} (z+d/2)\right] , \quad u_{R} = C_{R} \exp \left[ -\kappa_{R} (z-d/2)\right] .
\]
The spatial damping decrements   
\begin{equation}
\label{kapa_LR}
\kappa_{j} = \sqrt{ \frac{2m }{\hbar^{2} } \left( V_{j} - \frac{\hbar^{2} k_{z}^{2}}{2m} - \frac{V_{j} - 2E_{b}(\mathbf{k}_{\perp},k_{z}) }{2mc^{2}} V_{j} \right) } = \sqrt{(\kappa^{(0)}_{j})^{2} \left[1 - \nu_{j} + 2\mathit{e}_{b}(\bm{k}) \right] - k_{z}^{2}}   
\end{equation}
have the dimension of the wavevector. We have used here the following notations:
\begin{equation}
\label{kappa_0} 
\kappa^{(0)}_{j} = \sqrt{\frac{2m V_{j} }{\hbar^{2} }} ,  \quad \nu_{j} = \frac{V_{j}}{2mc^{2}} , \quad \mathit{e}_{b}(\bm{k}) = \frac{E_{b}(\mathbf{k}_{\perp},k_{z})}{2mc^{2}} ,
\end{equation}
where $ \nu_{j} $ ($ j = L,R $) and $ \mathit{e}_{b} (\bm{k})$ are dimensionless. In real situations the external electric field is applied to regions that exceed significantly dimensions of the QW. Therefore, strictly speaking, potentials depend on $ z $-coordinate also in regions beyond the QW, but in not too strong fields this dependence does not  practically affect the decreasing decrement (\ref{kapa_LR}).

In the central region the equation for function  $ u_{\sigma}(z) $ takes the form:
\[
-\hbar^{2} c^{2} \frac{d^{2} u_{\sigma}}{dz^{2}} - i\sigma \hbar c \frac{dV_{C}(z)}{dz} v_{\sigma} + \left[ 2E - V_{C}(z) \right] V_{C}(z) u_{\sigma} = \left[ E^{2} - \varepsilon^{2}\left( \mathbf{k}_{\perp} \right) \right] u_{\sigma} .
\]
In this equation at the non-relativistic energies (\ref{E_nonrel}) the denominator of (\ref{varphi}) can be approximated (cp. (\ref{E_nonrel})) as $ E + \varepsilon (\mathbf{K}_{\perp}) - V_{C}(z) \simeq 2E - V_{C}(z) \simeq 2mc^{2} $. Taking this into account and substituting the result into the equation, the latter is transformed to  the following form:
\begin{equation}
\label{Eq-psi_C} 
-\hbar^{2} c^{2} \left( \frac{d^{2} u_{\sigma}}{dz^{2}} - \frac{\mathcal{E}}{2mc^{2}} \frac{d u_{\sigma}}{dz} \right)  - 2mc^{2} \mathcal{E} z u_{\sigma} = \left[ E^{2} - \left(  \varepsilon^{2}\left( \mathbf{k}_{\perp} \right) - \sigma \frac{\hbar^{2} \mathcal{E} q}{2m} \right) \right] u_{\sigma} .
\end{equation}
The evident substitution  
\begin{equation}
\label{psi_C} 
u_{\sigma} \left( z \right) = e^{\kappa_{\mathcal{E}} z} w_{\sigma}(z)  
\end{equation}
with the field exponent
\begin{equation}
\label{lambda} 
\kappa_{\mathcal{E}} = \frac{\mathcal{E}}{4mc^{2}} ,
\end{equation}
reduces Eq. (\ref{Eq-psi_C}) to the SE-type equation:
\begin{equation}
\label{Eq-f_0} 
2mc^{2} \left( - \frac{\hbar^{2}}{2m} \frac{d^{2} w_{\sigma}}{dz^{2}} - \mathcal{E} z w_{\sigma} \right) = \left[ E^{2} - \left(  \varepsilon^{2}\left( \mathbf{k}_{\perp} \right) + \hbar^{2} c^{2}\kappa_{\mathcal{E}}^{2} - 2 \sigma \hbar^{2} c^{2} \kappa_{\mathcal{E}} q \right) \right] w_{\sigma}. 
\end{equation}
In this equation for the bound states with nonrelativistic energies there is a positive multiplier, which can be written as  
\[
E^{2} - \left[ \varepsilon^{2}\left( \mathbf{k}_{\perp} \right) + \hbar^{2} c^{2}\kappa_{\mathcal{E}}^{2} - 2 \sigma \hbar^{2} c^{2} \kappa_{\mathcal{E}} q \right]  = \hbar^{2} c^{2} k_{z}^{2},
\]
where $ k_{z} $ is some real parameter, which determines the eigen energy (see Eq. (\ref{E_nonrel})) 
\begin{eqnarray}
\label{E(q)} 
E = \sqrt{\varepsilon^{2}\left( \mathbf{k}_{\perp} \right) + \hbar^{2} c^{2}\kappa_{\mathcal{E}}^{2} - 2 \sigma \hbar^{2} c^{2} \kappa_{\mathcal{E}} q + \hbar^{2} c^{2}k_{z}^{2} } = \nonumber \\ 
= \sqrt{m^{2} c^{4} +  \hbar^{2}c^{2}\left( k_{\perp}^{2} + \kappa_{\mathcal{E}}^{2} + k_{z}^{2} - 2 \sigma \kappa_{\mathcal{E}} q \right)} \equiv mc^{2} + E_{b}(\mathbf{k}_{\perp},k_{z}).  
\end{eqnarray} 
As it will be seen below, the parameter $ k_{z} $ takes the discrete values $ k_{z} = k_{z}(n;\sigma , \mathbf{k}_{\perp}) $ and for its every value $ E_{b} = E_{n\sigma}(\mathbf{k}_{\perp}) $ represents the continuous band in the $k_{x}k_{y}$-plane. 

According to (\ref{Eq-f_0}), the function $ w_{\sigma} (z) $ satisfies the equation
\[
-\frac{\hbar^{2}}{2m} \frac{d^{2} w_{\sigma}}{dz^{2}}-\mathcal{E} zw_{\sigma} = \frac{\hbar^{2}k_{z}^{2}}{2m} w_{\sigma}, 
\]
which admits the exact solution  expressed via Airy functions \cite{Landau}. The analysis of the obtained results can be done numerically. Nevertheless, the qualitative  analysis can be also carried out  considering the potential energy $ \mathcal{E} z $ of the external electric field as a perturbation. The solution in the zero order approximation  is expressed via exponents $ \exp \left( \pm ik_{z} z \right) $ and the corrections to it can be obtained using the expansion: $ w(z) = \exp \left( ik_{z} z \right) \left( 1 + w_{1} + \ldots \right) $. The first order correction $ w_{1} $ is the solution of the equation 
\[
-\frac{\hbar^{2}}{2m} \frac{d^{2} w_{1}}{dz^{2}} - i\frac{\hbar^{2} k_{z}}{m} \frac{dw_{1}}{dz} = \mathcal{E} z .
\]
Therefore, in the first approximation the solution of Eq. (\ref{Eq-f_0}) is a linear combination  
\[w_{\sigma} (z) = A e^{ik_{z}z} g(z) + B e^{-ik_{z}z} g^{\ast}(z) ,\]
in which 
\begin{equation}
\label{g} 
g(z) = 1 + w_{1}(z) = 1 - \frac{m\mathcal{E} }{2\hbar^{2}k_{z}^{2}}z + i \frac{m\mathcal{E} }{2\hbar^{2}k_{z}} z^{2}.
\end{equation}
This solution is obviously valid if the inequality  $ (1/4)\mathcal{E}d \ll \hbar^{2}k_{z}^{2}/2m $ takes place. Taking into account Eq. (\ref{psi_C}), we can represent the function $ u_{\sigma} $ in the form 
\begin{equation}
\label{sol_psi} 
u_{\sigma} (z) = \left\lbrace \begin{array}{ll}
C_{L} e^{\kappa_{L} (z+d/2)}, &   z< -d/2 , \\
\left( A e^{ik_{z}z}g(z) + B e^{-ik_{z}z}g^{\ast}(z)\right) e^{\kappa_{\mathcal{E}} z}, &    -d/2 < z< d/2  , \\
C_{R} e^{-\kappa_{R} (z-d/2)}, &  z> d/2. 
\end{array} \right.  
\end{equation}

The small component $ v_{\sigma} $ in the corresponding regions is determined by Eq. (\ref{varphi}), which after the simple calculations yields
\begin{equation}
\label{sol_varphi}
 v_{\sigma} (z) = \left\lbrace \begin{array}{ll}
iC_{L}\frac{\sigma \hbar c}{E + \varepsilon (\mathbf{K}_{\perp})} \frac{  \sigma q - \kappa_{L}}{1 - \nu_{L} }  e^{\kappa_{L} (z+d/2)} , & z< -d/2  ,\\
i\frac{\sigma \hbar c}{E + \varepsilon (\mathbf{K}_{\perp})} \frac{f_{\sigma}(z)}{1 + 2 \kappa_{\mathcal{E}} z} ,  & -d/2 < z < d/2 , \\
iC_{R}\frac{\sigma \hbar c}{E + \varepsilon (\mathbf{K}_{\perp})}\frac{ \sigma q + \kappa_{R}}{1 - \nu_{R} }  e^{-\kappa_{R} (z-d/2)} , & z> d/2 ,
\end{array} \right.  
\end{equation}
where  
\begin{equation}
\label{upsilon} 
\begin{array}{c}
f_{\sigma}\left( z \right) = A \left[ \left( \sigma q - \kappa_{\mathcal{E}} - ik_{z} \right) g(z) - dg/dz \right] e^{\left( \kappa_{\mathcal{E}} + ik_{z} \right)z } + \\ + B \left[ \left( \sigma q - \kappa_{\mathcal{E}} + ik_{z} \right) g^{\ast}(z) - dg^{\ast}/dz \right] e^{\left( \kappa_{\mathcal{E}} - ik_{z} \right)z } .
\end{array} 
\end{equation} 
Here we have again introduced parameters $ \nu_{j} = V_{j}/\left( E + \varepsilon (\mathbf{K}_{\perp}) \right) \simeq V_{j}/2mc^{2}  $ ($ j= L,R $), which reduce to expressions (\ref{kappa_0}) if one neglects  the higher order relativistic corrections. 

The coefficients $ A $, $ B $, $ C_{L} $ and $ C_{R} $, and the admitted values of the parameter $ k_{z} $ can be determined from the boundary condition of the function (\ref{bcond}) and from the normalization. The equality (\ref{bcond}) for bispinors can be reduced to the matching condition for the functions  $ u_ {\sigma} $ and $ v_{\sigma} $ in the corresponding points. Obtaining coefficients $ C_{L} $ and $ C_{R} $ from this  condition for the functions  $ u_{\sigma} $, from the continuity condition  of the functions $ v_{\sigma} $ we get the system of two homogeneous  equations for coefficients  $ A $ and $ B $: 
\begin{equation}
\label{eqsAB}  
\left\lbrace \begin{array}{c}
F_{R} e^{ik_{z}d/2} A + F_{R}^{\ast} e^{-ik_{z}d/2} B = 0 , \\
F_{L}^{\ast} e^{-ik_{z}d/2} A + F_{L} e^{ik_{z}d/2} B = 0 \, ,
\end{array} \right. 
\end{equation}
where 
\begin{equation}
\label{F_LR} 
\begin{array}{c}
F_{R} = \left( \frac{\kappa_{R}}{1-\nu_{R}} + \frac{\kappa_{\mathcal{E}}}{1-\kappa_{\mathcal{E}}d} + \frac{\sigma \left( \nu_{R} - \kappa_{\mathcal{E}}d\right)q }{(1-\nu_{R})(1- \kappa_{\mathcal{E}}d )} + i\frac{k_{z}}{1-\kappa_{\mathcal{E}}d}  \right) g\left( d/2 \right) + \frac{1}{1-\kappa_{\mathcal{E}}d} \frac{dg}{dz}|_{z=d/2}, \\
F_{L} = \left(\frac{\kappa_{L}}{1-\nu_{L}} - \frac{\kappa_{\mathcal{E}}}{1+\kappa_{\mathcal{E}}d} - \frac{\sigma q \left( \nu_{L} + \kappa_{\mathcal{E}}d\right) }{(1-\nu_{L})(1+ \kappa_{\mathcal{E}}d )} + i\frac{k_{z}}{1+\kappa_{\mathcal{E}}d} \right) g^{\ast} \left( - d/2 \right) -  
\frac{1}{1+\kappa_{\mathcal{E}}d} \frac{dg^{\ast}}{dz}|_{z=-d/2},
\end{array}
\end{equation}
and the function $g(z)$ is determined in (\ref{g}). 

Using now the polar form for complex coefficients (\ref{F_LR})
\[
F_{j} = \mathrm{Re} F_{j} + i \mathrm{Im} F_{j} = \vert F_{j} \vert e^{i\Theta_{j} } , \quad \vert F_{j} \vert = \sqrt{\left(\mathrm{Re} F_{j} \right)^{2} + \left(\mathrm{Im} F_{j} \right)^{2} } , \quad \tan \Theta_{j} = \frac{\mathrm{Im}F_{j}}{\mathrm{Re}F_{j}}, 
\]
we see that the condition of the existence of a non-trivial solution of the system (\ref{eqsAB}),
\[ F_{L} F_{R} e^{ik_{z}d}- F_{L}^{\ast} F_{R}^{\ast} e^{-ik_{z}d}  = 0  ,\]
requires the fulfilment of the equality 
\[ \sin \left( k_{z}d + \Theta_{L} + \Theta_{R} \right) = 0 ,  \]
from which it directly follows that 
\begin{equation}
\label{cond_main} 
k_{z}d + \Theta_{L} + \Theta_{R} = \pi n , \quad n = 1, 2, 3, \ldots .
\end{equation}
Here the angles $ \Theta_{j} $ are nothing else as functions $ k_{z} $. With account of their determinations, the condition (\ref{cond_main}) can be rewritten as 
\begin{equation}
\label{cond1} 
k_{z}d + \arcsin \frac{\mathrm{Im} F_{R}}{\vert F_{R} \vert}  + \arcsin \frac{\mathrm{Im} F_{L}}{\vert F_{L} \vert} = \pi n .
\end{equation}

The allowed  values of the parameter $ k_{z} $ correspond to the roots $ k_{z}(n) $ of the transcendent Eq.  (\ref{cond1}) and determine  the discrete energies (see Eq. (\ref{E(q)})) of the bound electrons/holes (particles confined by the QW) with $ k_{z} = k_{z}(n) $. These energies increase with $ n $ increasing up to some value $ n_{max} $ which depends on the QW form.  Equation (\ref{cond_main}) contains angles $ \Theta_{L} $ and $ \Theta_{R} $, that explicitly, via functions (\ref{F_LR}), depend on spin number $ \sigma $ and parameter $ q $. The latter is directly (according to Eq. (\ref{q})) connected with the wavevector $ \mathbf{k}_{\perp} $. Thus, the roots of the equation are defined by the spin state, which, in its turn, affects the energy spectrum of 2D electron bands. Therefore, we set $ k_{z} = k_{z}(n;\sigma , \mathbf{k}_{\perp}) $ in Eq. (\ref{E(q)}). Respectively, the energy spectrum of electrons in QWs,  $ E_{b}(\mathbf{k},k_{z}) $, according to Eq.(\ref{E(q)}), takes the standard nonrelativistic form 
\begin{equation}
\label{calE}
E_{b}(\mathbf{k}_{\perp},k_{z}) \equiv E_{n \sigma}(\mathbf{k}_{\perp}) =  \frac{\hbar^{2}}{2m} \left[ k_{\perp}^{2} + \kappa_{\mathcal{E}}^{2} + k_{z}^{2}(n;\sigma , \mathbf{k}_{\perp}) - 2 \sigma \kappa_{\mathcal{E}} q \right]  .
\end{equation}

In general, Eqs. (\ref{cond1}) cannot be solved analytically and require numerical or graphical solutions. Examples of this equation graphical solution  for the lowest possible bound state $ n=1 $ at $ \mathcal{E} = 0 $ are given in Fig. 1. In this case Eq. (\ref{cond1}) becomes 
\begin{equation}
\label{cond_1-0} 
k_{z}d + \arcsin \dfrac{k_{z}}{\sqrt{\left(\frac{\kappa_{R} + \sigma \nu_{R} q }{1-\nu_{R}} \right)^{2} + k_{z}^{2} }} + \arcsin \dfrac{k_{z}}{\sqrt{\left( \frac{\kappa_{L} - \sigma \nu_{L} q}{1-\nu_{L}} \right)^{2} + k_{z}^{2} }} = \pi .
\end{equation}
As it is seen, it has the form $ F_{\sigma}(k_{z}) = \pi $, where the function $ F_{\pm}(k_{z}) $ is the left hand side of Eq.(\ref{cond_1-0}). Therefore, the solutions correspond to the intersection of the  curves $ y= F_{\sigma}(k_{z}) $ with the line $ y= \pi $. Since we assume $ V_{L} \geq V_{R} $, then according to Eq. (\ref{kapa_LR}) the variable $ k_{z} $ takes its values in the interval $ 0 \leq k_{z}^{2} \leq \left(\kappa^{(0)}_{R}\right)^{2} \left[ 1 - \nu_{R} + 2 \mathit{e}_{b}(k) \right] $.

Let us introduce the energy 
\begin{equation}
\label{E_0}
E_{0} = \frac{\hbar^{2}\pi^{2}}{2md^{2}} ,
\end{equation}
which is determined by the width of the QW for a particle with the given mass and defines the energy of the ground state (zero-point energy) of a particle in the infinitely deep QW (when the lowest solution corresponds to the wavevector $ k_{z}^{(0)} = \pi/d $). The energy $ E_{0} $ and potential heights $ V_{L} $ and $ V_{R} $  characterize completely the potential layer. For subsequent calculations it is convenient to introduce dimensionless parameters:
\begin{equation}
\label{tau-ksi} 
\tau = \sqrt{\frac{V_{R}}{E_{0}}} = \frac{\kappa^{(0)}_{R} d}{\pi} , \qquad \xi = \sqrt{\frac{V_{R}}{V_{L}}} = \frac{\kappa^{(0)}_{R}}{\kappa^{(0)}_{L}} \leq 1 , \qquad e_b^{(0)} = \frac{E_{0}}{2mc^{2}} .
\end{equation}
The parameter $ \tau $, which is introduced as the ratio of the lowest potential wall $ V_{R} $ to the characteristic energy, defines, in fact, the real depth of the QW. At $ \tau \leq 1 $ not more than one bound state can exist in the QW. The parameter $ \xi $ characterizes the QW asymmetry and $ \xi \leq 1 $ because $ V_{L} \geq V_{R} $. The value $ \xi = 1 $, obviously, corresponds to a symmetric QW in the absence of an external field. The parameter $ e_b^{(0)} \ll 1 $  (cp. (\ref{kappa_0})) determines relativistic corrections, as it will be shown below.

The corresponding solutions for two different values of the parameter $ \tau $ are presented in Fig.1 where the variable $ k_{z}d $ takes its values in the interval $ 0 \leq k_{z} d \leq \pi \tau $.
As it follows from the definition (\ref{q}), parameter $ q $ is the product of two values. Therefore, the curves $ F_{\pm}(k_{z}) $ split at any finite value $ \mathbf{k}_{\perp} $ under the condition  $ f(\varphi_{\bot}) \neq 0 $, and, hence, different spin states correspond to different roots of  Eq.(\ref{cond1}), what is the direct indication that the bound state energy depends on the spin quantum number, indeed.

%\vspace{1cm}
\begin{figure}[t]
\begin{picture}(16,8)
  \includegraphics[width=5cm]{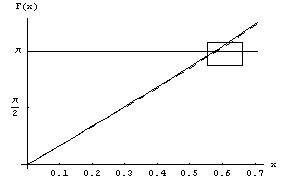}
   \includegraphics[width=5cm]{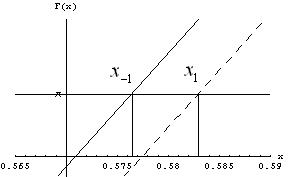}
  \includegraphics[width=5cm]{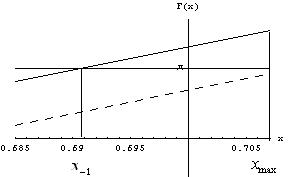}
  \end{picture}
  \caption{Solution of the equation  $ F(x)=\pi $ at $ \xi = 0.5$, $\nu_0 =10^{-2}$   a) in the whole interval of the variable $x=k_zd$, $0<x<\pi \tau $ at $\tau = 2.5$; b)  in the interval of $x$ close to the roots of the solution; c) solution at $\tau =0.5$, when there is only one root of the equation  $ F(x)=\pi $, denoted as $x_{-1}$. Here $F$ is the left hand side of Eq. (\ref{cond_1-0}).  }
  \label{fig:1}
\end{figure}
 
In the interval $ 0 \leq k_{z} d \leq \pi \tau $ the functions $ F_{\pm}(k_{z}) $ increase from $ 0 $ to $ F_{\pm}(\pi \tau) $. For the existence of the roots it is necessary that the inequality $ F_{\pm}(\pi \tau) \geq \pi $ is fulfilled which is the condition for the existence of at least one bound state. As it is well known, in a shallow asymmetric potential box the inequality can take place $ F_{\pm}(\pi \tau) < \pi $ at $ \tau $ less than some critical value which depends on the asymmetry parameter $ \xi $. It is important that at  finite values $ \mathbf{k}_{\perp} $ and at nonzero value $ q $, the  critical values of the parameter $ \tau $ are different for different spin states. This means that for some parameter values of the QW and wavevectors $ \mathbf{k}_{\perp} $, the equation (\ref{cond1}) can possess the root for the  only value of $ \sigma $, and doesn't possess the root for  $- \sigma $. In other words, the situation is possible when a QW captures electrons with one spin projection only, leaving electrons with opposite projection free. Such example is shown in Fig. 1c.

For qualitative analysis of dispersion laws (\ref{calE}) let us consider the approximate solutions of Eq.(\ref{cond1}) for the lowest bound state, which corresponds to  $ n=1 $, in two different cases: \textit{ i)} enough deep QW and \textit{ii)} strongly asymmetric shallow QW.

\textit{i)} \textbf{Deep QW.} In this case when several bound states exist, $ \tau^{2} \gg 1 $, the lowest energy for the particle is much smaller than the potential wall heights. So, the inequality $ k_{z}^{2}(1)/ \left( \kappa^{(0)}_{j} \right)^{2} = E_{1} / V_{j} \ll 1 $ ($ E_{1} = \hbar^{2}k_{z}^{2}(1)/2m $) is fulfilled. Taking into account that $ \mathrm{Im} F_{j} \sim k_{z} $ and $ \mathrm{Re} F_{j} \sim \kappa^{(0)}_{j} $, the arguments of trigonometric functions in Eq.(\ref{cond1}) are small, one can use the Taylor expansion. This will reduce the equation to the form 
\[
k_{z} d = \pi - X(k_{z}) , \quad X(k_{z}) \approx \frac{\mathrm{Im} F_{R}\,  \mathrm{Re} F_{L}  + \mathrm{Im} F_{L}\, \mathrm{Re} F_{R}  }{\mathrm{Re} F_{L}\, \mathrm{Re} F_{R} } \equiv \Delta \left(k_{z} \right) k_{z}d .
\]
The method of successive approximations can be used in view of the smallness of $ X(k_{z}) $. In the zero approximation we have $ k_{z}^{(0)} = \pi /d $. The first order approximation is  
\[
k_{z}^{(1)} =k_{z}^{(0)}-X(k_{z}^{(0)})/d
 =  \left( 1 - \Delta \right) k_{z}^{(0)} , 
\]
where $\Delta$ is a small correction of the following form
\begin{equation}
\label{Delta} 
\begin{array}{c}
\Delta = \Delta \left(k_{z}^{(0)} \right) = \Delta_{0} - \sigma \Delta_{1} qd + \Delta_{2} q^{2}d^{2} ,  \\  
\Delta_{0} = \frac{\left(\sqrt{V_{L}} + \sqrt{V_{R}} \right)\sqrt{E_{0}} }{\pi \sqrt{V_{L} V_{R}}} + \frac{\sqrt{V_{L}} - \sqrt{V_{R}}}{4\pi \sqrt{V_{L}V_{R}E_{0}}} \mathcal{E}d + \left( \frac{\mathcal{E} d}{8E_{0}} \right)^{2} ,  \\
\Delta_{1} = \left[ \frac{\left( V_{L}-V_{R} \right)E_{0} }{\pi^{2} V_{L}V_{R}} + \frac{\mathcal{E} d}{2E_{0}} \right] \frac{E_{0}}{2mc^{2}} , \\
\Delta_{2} = \left[ \frac{\left(\sqrt{V_{L}} + \sqrt{V_{R}} \right) \left( 3\sqrt{V_{L}V_{R}} -V_{L} -V_{R} \right)}{\pi^{3}\sqrt{V_{L}V_{R}E_{0}}} - \frac{\left(V_{L}^{1/2} - V_{R}^{1/2} \right) \left(V_{L}^{3/2} + V_{R}^{3/2} \right) }{4\pi^{3} \left(V_{L}^{1/2} + V_{R}^{1/2} \right) \sqrt{V_{L}V_{R}E_{0}}} \frac{\mathcal{E} d}{E_{0}} \right] \left( \frac{E_{0}}{2mc^{2}} \right)^{2}. 
\end{array}
\end{equation}
To obtain this solution, we have used the explicit expressions (\ref{kapa_LR}), (\ref{kappa_0}),  (\ref{g}), (\ref{F_LR}), keeping main terms only accounting for the smallness of the value  $ 1/\pi \tau $, the smallness of the ratio of nonrelativistic energies to the energy $ mc^{2} $, and the smallness of the ratio  $ \mathcal{E}d /E_{0} $. Notice, the approximate solution contains relativistic corrections and includes the terms, which are determined by the QW form and by the strength of the field. 

Substituting the first approximation $ k_{z} \simeq k_{z}^{(1)} $ in Eq. (\ref{calE}) and using the  definitions (\ref{q}) and (\ref{Delta}), for the lowest 2D band of electrons in a QW we find Rashba-like dispersion law :
\begin{equation}
\label{calE_2D} 
E_{1 \sigma}(\mathbf{k}_{\perp})  = E_{1}\left( 0 \right)  + \frac{\hbar^{2} k_{\perp}^{2}}{2m^{\ast}} + \sigma \alpha_{BR} k_{\perp} ,
\end{equation}
where 
\begin{equation}
\label{calE_0}
E_{1}\left( 0 \right) = \left( 1 - 2 \frac{1 + \xi}{\pi \tau } \right) E_{0} - \frac{1-\xi } {2\pi \tau } \mathcal{E}d - \frac{\left( \mathcal{E}d \right)^{2} }{32 E_{0}} 
\end{equation}
is the energy at $ k_{\perp} = 0 $ with account of the QW depth and external field. The value
\begin{equation}
\label{massa} 
\begin{array}{c}
m^{\ast} \equiv m^{\ast}_{deep}\left( \varphi_{\bot} \right) = m \left\{ 1 -M\left[ e_b^{(0)}  f(\varphi_{\bot})\right]^2 \right\}^{-1} , \\ 
M = \frac{2\tau}{\pi \xi^{2}} \left[\left( 1+\xi \right) \left(1-3\xi +\xi^{2} \right)  + \left( 1-\xi \right) \left(1-\xi +\xi^{2} \right) \frac{\mathcal{E}d}{4E_{0}} \right] 
\end{array}
\end{equation}
is the effective mass renormalized by SOI. The coefficient  
\begin{equation}
\label{alpha_BR} 
\alpha_{BR} \equiv \alpha_{BR}^{deep}\left(\varphi_{\perp} \right) = \frac{2e_b^{(0)} d}{\pi^{2}} \left( \frac{1-\xi^{2}}{\tau^{2}} E_{0} + \frac{\pi^{2} - 1}{2} \mathcal{E}d \right) f\left( \varphi_{\perp} \right) 
\end{equation}
is usually called the Bychkov-Rashba parameter. In general, it as well as the effective mass (\ref{massa}) can be anisotropic due to the presence of function $ f\left( \varphi_{\perp} \right) $. It is easy to see that the values of $ m^{\ast} $ and $ \alpha_{BR} $ are determined by the internal structural asymmetry of the QW and by the applied external field.

\textit{ii)} \textbf{ Strongly asymmetric shallow QW}. Here the parameter $ \tau < 1 $ and no more than one bound state can exist, the assumptions used to solve Eq.(\ref{Eq-f_0}) can be fulfilled in very weak fields only. Therefore we restrict consideration to the case when the field is absent.  Then the Eq.(\ref{cond_1-0}) can be represented in the form  
\begin{equation}
\label{cond3_x} 
k_{z}d + \arctan \frac{\left(1-\nu_{R} \right)k_{z} }{ \kappa_{R} + \sigma \nu_{R} q } + 
\arctan \frac{\left(1-\nu_{L} \right)k_{z} }{ \kappa_{L} - \sigma \nu_{L} q } = \pi .
\end{equation}

The roots of Eq.(\ref{cond_1-0}) or Eq.(\ref{cond3_x}) for shallow QW correspond to the values of $ k_{z} $ that are close to $ \kappa_{R}^{(0)} $ such that the inequality $ \left( \kappa_{R} / \kappa_{R}^{(0)} \right)^{2} \ll 1 $ is valid. So, one can write $ k_{z}^{2} = \left( \kappa_{R}^{(0)} \right)^{2} - \kappa_{R}^{2} \approx \left( \kappa_{R}^{(0)} \right)^{2} $, and the argument of the first arctangent in the right hand side of Eq. (\ref{cond3_x}) becomes large. For strongly asymmetric QW the inequality $ k_{z}^{2}/\kappa_{L}^{2} \sim \left( \kappa_{R}^{(0)}/\kappa_{L}^{(0)} \right)^{2} = \xi^{2} \ll 1 $ is also true, and the argument of the second arctangent is small, as in the case of a deep QW. Using the expansions 
\[
\arctan y = \frac{y}{\sqrt{1 + y^{2}}} \left( 1 + \frac{1}{6} \frac{y^{2}}{1 + y^{2}} + \ldots \right), \quad y^{2} \ll 1  ,
\]
\[
\arctan y = \frac{\pi}{2} - \frac{1}{y} + \frac{1}{3y^{3}} - \ldots, \quad  y^{2} \gg 1,  
\]
one can reduce Eq. (\ref{cond3_x}) in the first approximation to the form 
\[
k_{z}d - \frac{\kappa_{R} + \sigma \nu_{R}q}{k_{z}} + \frac{k_{z}}{\sqrt{\left( \kappa_{L} - \sigma \nu_{L} q \right)^{2} + k_{z}^{2}}} = \frac{\pi}{2} .
\]
Keeping the main terms only, from this relation we obtain that 
\begin{equation}
\label{r_sol} 
\kappa_{R} = \kappa_{R}^{(0)} \left( a_{0} + \sigma a_{1} q  + a_{2} q^{2} \right),  
\end{equation}
where coefficients $ a_{0} $, $ a_{1} $ and $ a_{2} $ are expressed via dimensionless parameters (\ref{tau-ksi}):
\begin{equation}
\label{a_123} 
\begin{array}{c}
a_{0} = \pi \left( \tau - \tau_{thr} \right) , \quad  a_{1} = \xi^{2} \frac{\sqrt{E_{0} V_{R}}}{2\pi mc^{2}} , \quad a_{2} = \xi \left( \frac{\sqrt{E_{0} V_{L}}}{2\pi mc^{2}} \right)^{2}, \\
\tau_{thr} = \frac{\pi}{2} - \xi \left(1+\xi^{2} \right). 
\end{array}
\end{equation}
It follows from (\ref{r_sol}) that the roots of Eq. (\ref{cond3_x}) exist when the right hand side of Eq. (\ref{r_sol}) is positive. In the case when $q=0$, this means that  
 electrons can be confined by the asymmetric shallow QW if 
$
\tau > \tau_{thr} $.
When $q \neq 0$, the threshold values of the parameter $\tau $ are different for different quantum spin numbers $\sigma$. In the vicinity of the threshold,    $ \tau \approx  \tau_{thr} $,  the bound state corresponds to one value of the spin quantum number only, namely, to the value  $ \sigma = 1 $. 

At $ \pi \left( \tau - \tau_{thr} \right) > a_{1}q $, there exist bound states for both values of the spin projection and the solution follows directly from Eq. (\ref{r_sol}):  
\[ 
k_{z}^{2} = \left( \kappa_{R}^{(0)} \right)^{2} \left[ 1 - a_0^{2} - 2\sigma a_{0} a_{1} qd - \left( a_{1}^{2} + 2a_{0}a_{2} \right) q^{2}d^{2} \right] .
\]
Substituting this result in Eq. (\ref{calE}) and using definitions (\ref{a_123}), we find the expression for 2D electron band (\ref{calE_2D}), in which
\begin{equation}
\label{calE_0_s} 
E_{1}\left( 0 \right) = V_{R}\left[  1 - \pi^{2} \left( \tau - \tau_{thr} \right)^{2} \right], 
\end{equation} 
the effective mass is
\begin{equation}
\label{m_sh} 
m^{\ast} \equiv m^{\ast}_{shallow}\left( \varphi_{\bot} \right) = m \left\{1 -\left( \xi^{4} + 2\pi \frac{\tau - \tau_{thr}}{\xi} \right)\left[e_b^{(0)}  \tau  f(\varphi_{\bot})\right]^2 \right\}^{-1} , 
\end{equation}
and the Bychkov-Rashba parameter is
\begin{equation}
\label{alpha_R_s} 
\alpha_{BR}\equiv \alpha_{BR}^{shallow}(\varphi_{\bot}) =  - 2 e_b^{(0)}   \xi^{2} \tau \left( \tau - \tau_{thr} \right) f(\varphi_{\bot})V_{R} d .
\end{equation}

Finally, comparison of Bychkov-Rashba parameters for deep and shallow QWs, shows that in the former case it increases with $\tau $ decreasing, while in the latter case it decreases. This means that  Rashba spin splitting can have maximum  at  $ \tau \sim 1$.  This situation   depends on the specific geometry of QW, and, thus, can be controlled via synthesis of the corresponding spintronics heterostructures.

\section{Conclusion} 

The found above general solution of the DE, is important for deep understanding of spin and spin-related properties of various systems. In particular, the main characteristic of such systems from the point of view of spintronics applications, is spin orientation which is characterized by the vector $ \mathbf{S}_{\sigma}(z) = \left\langle \varPsi_{\sigma} \vert \bm{\hat{\Sigma}} \vert \varPsi_{\sigma} \right\rangle $. For solutions with positive energy, the lower spinor, according to Eqs. (\ref{psi_u})-(\ref{psi_d}), always is a small component of the bispinor, although both small and large components can be present in all four components of the bispinor.  Therefore, neglecting relativistic corrections,  $ \mathbf{S}_{\sigma} $  can be approximated by the mean value of the Pauli vector matrices over the upper spinor of the bispinor $ \varPsi_{\sigma} $, namely  $ \mathbf{S}_{\sigma}(z) \approx \left\langle \psi_{\sigma, u} \vert \bm{\hat{\sigma}} \vert \psi_{\sigma, u} \right\rangle $. Moreover, it is sufficient to take into account the large component of the solution only; then, according to  Eq. (\ref{psi_u}), we have $ \mathbf{S}_{\sigma}(z) = \vert u_{\sigma}(z) \vert^{2} \mathbf{s}_{\sigma} $, where 
\begin{equation}
\label{s_pol} 
\langle \tilde{\chi}_{\sigma} \vert \bm{\hat{\sigma}} \vert \tilde{\chi}_{\sigma} \rangle \equiv \mathbf{s} = \sigma \left(  \mathbf{e}_{x} \sin \theta \cos \varphi + \mathbf{e}_{y} \sin \theta \sin \varphi + \mathbf{e}_{z} \cos \theta \right) .
\end{equation}
This vector, in its turn, is characterized by the spin quantum number, $ \sigma $, and, what is important,  depends on free parameters  $ \theta $ and $ \varphi $. Namely these parameters govern the spin degree of freedom as a physical characteristics of a particle.

In a homogeneous  space, the solution of the DE is a bispinor (\ref{Psi_free}) with arbitrary spinors  $ \chi $ and $ \eta $, whose components determine the spin, i.e., spin orientation.  An arbitrary spinor in a laboratory coordinate system has the form  (\ref{spinors}). Therefore, in a homogeneous  space the spin and spatial degrees of freedom of a particle are uncoupled: a particle can propagate in all directions  preserving  the  given (prepared "beforehand") spin orientation, which, generally speaking, can be arbitrary.  Consequently, in the nonrelativistic limit spin and spatial coordinates of a particle prove to be separated.  In other words, spin orientation is not manifested in energetic spectrum or other observables in the absence of external fields (electric and/or magnetic). 

The situation changes drastically when a particle moves in an inhomogeneous   potential. In particular, in the field which preserves free propagation of a particle in two directions and restricts it in the third one, the spin orientation determined by free parameters, can be arbitrary, according to the found above general solution, but 2D electron energy spectrum  
(\ref{calE_2D}) does depend on the spin. Such energy dependence on the spin variable reflects the manifestation of the SOI which appears in the SE. The specific form of the SOI is the consequence of the account of relativistic corrections. Even being small ones, they lead to qualitative consequences.

To demonstrate the possibilities of the description of qualitatively different characteristics of electrons in heterostructures with the help of general solution of the DE, let us show schematically spin characteristics for different non-coinciding solutions found in \cite{arXiv}. Figure 2 shows spin splitting and spin orientations in the $ k_{x}\,k_{y} $-plane at the constant value of the energy $ E_{\sigma}(\mathbf{k}_{\perp}) = E > E_1(0) $. In particular, Fig. 2a ($ \theta = \pi /2 $ and $ \varphi = \varphi_{\bot} + \pi/2 $) corresponds to the state with the given $z$-component of electric spin polarization $ \bm{\epsilon} $ providing the well-known Rashba spin-splitting. However,  choosing $ \varphi \equiv \varphi\left( \varphi_{\bot} \right)  = \varphi_{\bot} - \pi/2 $, we obtain the equivalent state, but with the opposite spin orientations in the sub-bands. Formally, this corresponds to the change of Bychkov-Rashba parameter sign. Figure 2b ($ \theta = 0 $) shows spin-degenerate electron spectrum (at zero value of Bychkov-Rashba parameter, $\alpha_{BR}=0$) when $z$-component of magnetic spin polarization, $ \bm{\mu}$,  is preserved.  Finally, in Fig. 2c ($ \theta = \pi /2 $, $ \varphi = 0 $) we show the case when the projection of spin pseudovector $ \bm{\mathcal{S}} $ is preserved (for instance, the projection on $x$-axis). In this case Bychkov-Rashba parameter becomes anisotropic, and, respectively, spin splitting becomes anisotropic too. 

 Figure 2 shows the simplest cases. Any arbitrary situation can be described by the general solution, in which one has to set certain values to the corresponding free parameters: for different physical situations certain {\it{a priori}} free parameters can be fixed. The essential benefit of the general and unique solution found in the present paper, is the possibility of continuous transition from one to another physical situation via continuous change of the corresponding free parameters. The resulting spin splitting and spin orientation schemes will have more complex geometry, than shown in Fig. 2. 

The above general solution describes the states of one isolated electron. 
Realization (“preparation”) of a specific spin state is, however, determined by another factors, such as carrier concentration, presence of electric or/and magnetic field(s), applied pressure, properties of particular interface etc. Therefore, spin states should be manifested as the peculiarities of real physical experiments.

In equilibrium systems the main requirement is the total energy minimum. For instance, for ideal 2D electron gas containing $ N_{e} $ electrons,  the total ground state energy is
\[
E_{tot} = \sum_{\sigma , \mathbf{k}_{\perp}} E_{\sigma}\left(\mathbf{k}_{\perp}\right) n_{\sigma \mathbf{k}_{\perp}}  , \quad \sum_{\sigma , \mathbf{k}_{\perp}} n_{\sigma \mathbf{k}_{\perp}} = N_{e} ,
\]
where $ n_{\sigma \mathbf{k}_{\perp}}  $ are occupation numbers of 
one-electron energy levels  (\ref{calE_2D}). At zero temperature $ n_{\sigma \mathbf{k}_{\perp}} =1 $  at $ E_{\sigma}\left(\mathbf{k}_{\perp}\right) \leqslant E_{F} $ and $ n_{\sigma \mathbf{k}_{\perp}} = 0 $ at $ E_{\sigma}\left(\mathbf{k}_{\perp}\right) > E_{F} $ with $ E_{F} $ being the Fermi energy. For the states, shown in Fig. 2, it is easy to find that $ E_{tot}^{\mu_{z}} > E_{tot}^{\mathcal{S}_{x}} > E_{tot}^{\epsilon_{z}} $ and 
\[
E_{tot}^{\mu_{z}} - E_{tot}^{\mathcal{S}_{x}} = N_{e} \frac{m \alpha_{BR}^{2}}{2\hbar^{2}} , \quad E_{tot}^{\mathcal{S}_{x}} - E_{tot}^{\epsilon_{z}} = \left\lbrace  \begin{array}{cc}
N_{e}\frac{\pi \hbar^{2} n_{e}}{2m} \left( 1-\frac{n_{e}}{3n_{0}} \right) , & n_{e} < n_{0}, \\
N_{e}\frac{\pi \hbar^{2} n_{0}}{2m} \left( 1-\frac{n_{0}}{3n_{e}} \right) , & n_{e} > n_{0}.
\end{array} \right. 
\]
Here $ n_{e} = N_{e}/S $ is the surface electron density,  $ n_{0} = m^{2}\alpha_{BR}^{2}/\pi \hbar^{4} $ is the maximum electron density at which the lowest spin sub-band is occupied and the  upper sub-band is unoccupied. The lowest spin sub-band has the minimum at $ k_{\perp} = k_{0} =\sqrt{ \pi n_0}$. 

Thus, in the equilibrium 2D electron gas, the Rashba state is realized as the lowest one. The presence of perturbations which destroy the isotropy of a 2D layer, can be taken into account by choosing certain values of the free parameters of the general solution, and, therefore,  one can hope that the situation  is not exclusive, when the lowest energy state is different from the Rashba state. 

\vspace{2cm}
\begin{figure}[t]
\begin{picture}(16,8)
  \includegraphics[width=5cm]{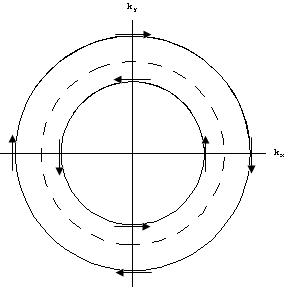}
   \includegraphics[width=5cm]{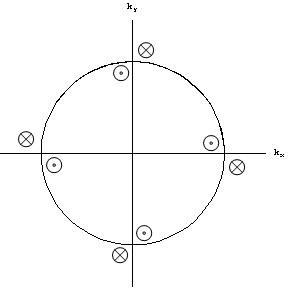}
  \includegraphics[width=5cm]{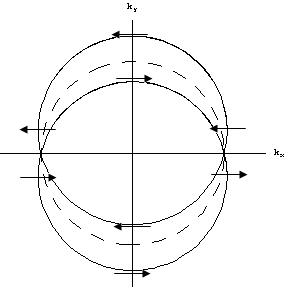} 
  \end{picture}
  \caption{Spin orientation schemes for three cases of preserved projections of spin polarization,  which follow from the general solution at the special choice of free parameters (see text): а) $\epsilon_z=const $,  b) $\mu_z=const$, c) $ \bm{\mathcal{S}}  \bm{e}_x =const$. }
  \label{fig:2}
\end{figure}

%\vspace{1cm}
{\textbf{Acknowledgement}} We express our sincere gratitude to E.I. Rashba for his fruitful comments. The work is done under the partial support of the Fundamental Research Program of the National Academy of Sciences of Ukraine.  

\appendix
\section{Solution of Dirac equation for free particles } 

The solution of DE for a free particle is well-known and can be found in textbooks (see, e.g.,  \cite{LandauIV}). The eigenfunctions of the DE are given by the bispinors 
\begin{equation}
\label{Psi_free} 
\Psi_{p ,\mathbf{k}}(\mathbf{r}) = A e^{i\mathbf{k}\cdot \mathbf{r}} \sqrt{\frac{\varepsilon + mc^{2}}{2\varepsilon }}
 { \chi \choose \frac{\hbar c\mathbf{k} \bm{\hat{\sigma}}}{\varepsilon + mc^{2}} \chi }  , \: \Psi_{a ,\mathbf{k}}(\mathbf{r}) = A e^{i\mathbf{k}\cdot \mathbf{r}} \sqrt{\frac{\varepsilon + mc^{2}}{2\varepsilon }}  
{ -\frac{\hbar c\mathbf{k} \bm{\hat{\sigma}}}{\varepsilon + mc^{2}} \eta \choose \eta } ,
\end{equation}
which describe the states of a particle with the momentum  $ \mathbf{p} = \hbar \mathbf{k} $, where $ \mathbf{k} = \mathbf{e}_{x} k_{x} + \mathbf{e}_{y} k_{y} + \mathbf{e}_{z} k_{z} $ is the corresponding wavevector.  In Eq. (\ref{Psi_free}) the notation $ \varepsilon \equiv \varepsilon (\mathbf{k}) = \sqrt{m^{2} c^{4} + \hbar^{2} c^{2} k^{2}} $ ($ k^{2} = k_{x}^{2} + k_{y}^{2} + k_{z}^{2} $) is used.   The first bispinor corresponds to the positive value of the eigen energy,  $ E = \varepsilon (\mathbf{k}) $, and the second one to the negative eigen energy, $ E = - \varepsilon (\mathbf{k}) $ (to particle  and antiparticle, respectively).  Notice, to have the complete system of eigen functions, it is necessary to supplement the bispinors (\ref{Psi_free}) by the  same bispinors  with Kramers partners of the  arbitrary spinors  $ \chi $ and $ \eta $. Two orthogonal bispinors can be characterized by the spin number $\sigma (=\pm 1)$ with opposite values. The simplest choice of such  spinors is in the frame of spin quantization axis:
\begin{equation}
\label{spinaxis} 
\chi_{+} =   \left( \begin{array}{c} 
1 \\ 
0
\end{array} \right) , \quad \chi_{-} = \left( \begin{array}{c} 
0 \\ 1
\end{array} \right) . 
\end{equation}
 Arbitrary spinors in (\ref{Psi_free}) with opposite spins can be expressed via the superposition of spinors (\ref{spinaxis}) in the form
\begin{equation}
\label{spinors} 
\tilde{\chi}_{+} =  e^{i\Theta /2} \left( \begin{array}{c} 
e^{-i\varphi /2} \cos \frac{\theta}{2} \\ 
e^{i\varphi /2} \sin  \frac{\theta}{2}
\end{array} \right) , \quad \tilde{\chi}_{-} = e^{-i\Theta /2} \left( \begin{array}{c} 
-e^{-i\varphi /2} \sin \frac{\theta}{2} \\ e^{i\varphi /2} \cos  \frac{\theta}{2}
\end{array} \right), 
\end{equation}
where $\Theta $ defines the general (global) phase of the wavefunction, and the angles $\theta $ and $\varphi $ are the polar angles that determine the particle spin direction  relative to the spin quantization axis. Here we have used the accent {\it{tilde }} to distinguish such arbitrary spinors from the ones associated with the eigen spin quantization axis (\ref{spinaxis}). 

Substituting (\ref{spinors}) in the function (\ref{Psi_free}) we obtain the eigen function $ \Psi_{\mathbf{k},\sigma }(\mathbf{r}; \Theta, \varphi, \theta) $ which is characterized by the quantum numbers determining particle motion, $ \mathbf{k} $, and its spin state, $ \sigma $, and which depends on the spatial and spin variables $ \mathbf{r} $ and $ \Theta , \, \varphi , \, \theta $, respectively. 

In the homogeneous space when $ V(\mathbf{r}) = 0 $, the transformation (\ref{U_trans}), which separates the wavevector $ \mathbf{k} $ into the components $ \mathbf{k}_{\bot} = \mathbf{e}_{x} k_{x} + \mathbf{e}_{y} k_{y} $ and $ \mathbf{e}_{z} k_{z} $, according to Section 3, is an intermediate step which somewhat complicates the calculations but does not affect the final result (\ref{Psi_free}). To show this, let us consider the solution of Eqs. (\ref{psi-phi}) at $ V(z) = 0 $ considering the states with positive energy only.

Excluding the function $ v_{\sigma}(z) $, we obtain Eq. (\ref{Eq_psi_LR}) for function $ u_{\sigma}(z) $ which admits the solution in the form of a wave  $ u_{\sigma} (z) = C \exp (ik_{z} z) $ with the eigenvalue 
\begin{equation}
\label{E_free} 
E = \sqrt{\varepsilon^{2} \left( \mathbf{k}_{\perp} \right)  + \hbar^{2} c^{2} k_{z}^{2} } = \sqrt{m^{2} c^{4} + \hbar^{2} c^{2} k^{2}} \equiv  \varepsilon (\mathbf{k}) , \; k^{2} = k_{\bot}^{2} + k_{z}^{2} .
\end{equation}
Determining the small component $ v_{\sigma} (z) $ from Eq. 
 (\ref{varphi}), one can obtain  
the eigen bispinors of the equation $ \hat{\tilde{H}} \tilde{\varPsi} = E \tilde{\varPsi} $  (cp. (\ref{sol_up-d})): 
\begin{equation}
\label{bispin_free} 
\tilde{\varPsi}_{k_{z},\sigma} (z) = C  \left( \begin{array}{c} 
\chi_{\sigma} \\ \frac{\hbar c \left( k_{z} \hat{\sigma}_{z} + iq \right) }{\varepsilon \left( \mathbf{k} \right) + \varepsilon \left( \mathbf{K}_{\perp} \right)} \chi_{\sigma}
\end{array} \right)e^{ik_{z}z},  
\end{equation}
where spinors $ \chi_{\sigma} $ are defined in Eq. (\ref{spinaxis}). The wavefunction $\tilde{\varPsi}_{k_{z},\sigma} (z)$  is  $\tilde{\varPsi}_{\sigma} (z)  = \hat{U}^{-1} \varPsi_{k_{z},\sigma} (z) $, where the operator $ \hat{U} $ is determined by the matrix (\ref{U_trans}). Using the explicit expressions for the transformation matrices (\ref{U_trans})-(\ref{U_pm}) and the definitions (\ref{K_x-K_y}) and (\ref{q}), one can show that the final wavefunction (\ref{Psi_2D}) takes the form of the solution (\ref{Psi_free}):
\begin{equation}
\label{Psi_our} 
\Psi_{\mathbf{k} , \sigma}(\mathbf{r}) = C \hat{U} \left( \begin{array}{c} 
\chi_{\sigma} \\ \frac{\hbar c \left( k_{z} \hat{\sigma}_{z} + iq \right) }{\varepsilon \left( \mathbf{k} \right) + \varepsilon \left(( \mathbf{K}_{\perp} \right)} \chi_{\sigma}
\end{array} \right)e^{i\mathbf{k} \mathbf{r}}  =
C  \sqrt{\frac{\varepsilon (\mathbf{k}) + mc^{2}}{2\varepsilon (\mathbf{k}) }} { \tilde{\chi}_{\sigma} \choose \frac{\hbar k \mathbf{e}_{\mathbf{k}} \bm{\hat{\sigma}} }{\sqrt{\left(mc \right)^{2} +\left( \hbar c\right)^{2} }}  \tilde{\chi}_{\sigma} } e^{i\mathbf{k}\cdot \mathbf{r}}
\end{equation}
with the spinor $ \tilde{\chi}_{\sigma} $, determined as $ \tilde{\chi}_{\sigma} = \hat{\omega} \chi_{\sigma} $ with $ \hat{\omega} = \hat{\omega}_{u} \hat{\omega}_{q} $. Here the rotation matrix  $ \hat{\omega}_{u} $ is defined in Eq. (\ref{U_pm}), and 
\begin{equation}
\label{omega_ell} 
\hat{\omega}_{q} = \left(  \begin{array}{cc} 
\cos \frac{\vartheta_{q}}{2} & -ie^{-i\varphi_{q}} \sin \frac{\vartheta_{q}}{2} \\ -ie^{i\varphi_{q}} \sin \frac{\vartheta_{q}}{2} & \cos \frac{\vartheta_{q}}{2} 
\end{array}  \right) , 
\end{equation}
where the angles $ \vartheta_{q} $ and $ \varphi_{q} $ are given by the equalities
\begin{equation}
\label{varteta} 
\tan \vartheta_{q} = \frac{\hbar^{2} c^{2} K_{\bot} \sqrt{k_{z}^{2} + q^{2}}}{mc^{2} \varepsilon \left( \mathbf{k} \right) + \varepsilon^{2} \left( \mathbf{K}_{\bot} \right) } , \quad 
\tan \varphi_{q} = \frac{q K_{y} - k_{z} K_{x}}{q K_{x} + k_{z} K_{y}} .
\end{equation}
Because the parameters in the matrix $ \hat{\omega}_{u} $ are arbitrary, they can be redefined  in such a way that their arbitrariness is completely assigned to the final matrix $ \hat{\omega} $ with the Keyli-Klein parameters (\ref{KK-Eil}). Then the transformation $ \tilde{\chi}_{\sigma} = \hat{\omega} \chi_{\sigma} $ leads to the spinors (\ref{spinors}) and, as a result, the wavefunction (\ref{Psi_our}) fully coincides with (\ref{Psi_free}).

\end{document}